\documentclass[aps,prd,showpacs,floatfix,notitlepage,nofootinbib,twocolumn,a4paper]{revtex4}
\usepackage{amsfonts,amsmath,units,wasysym,epsfig,graphicx,verbatim,color,subfigure,graphicx,bm,mathrsfs,lipsum,hyperref,cleveref}
\usepackage{booktabs}
\usepackage[normalem]{ulem}  
\definecolor{dgreen}{RGB}{0, 204, 0}

\newcommand{\be}{\begin{equation}}
\newcommand{\ee}{\end{equation}}
\newcommand{\bea}{\begin{eqnarray}}
\newcommand{\eea}{\end{eqnarray}}

\DeclareUnicodeCharacter{2212}{-}
\usepackage[none]{hyphenat}

\usepackage{cellspace}
\usepackage{appendix}

\begin{document}


\title{Joint Bayesian Inference of Dense Matter and \(f(Q)\) Gravity with Nuclear and Neutron Star Constraints}

\author{\href{https://orcid.org/0000-0003-0103-5590}N. K. Patra$^{1}$\includegraphics[scale=0.06]{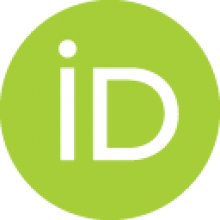}}
 \email{nareshkumarpatra@cuhk.edu.cn}
 
\author{\href{https://orcid.org/0000-0002-3223-4085}Sneha Pradhan$^{2,3}$\includegraphics[scale=0.06]{Orcid-ID.png}}
\email{snehapradhan2211@gmail.com}

 \author{\href{https://orcid.org/0000-0003-2130-8832}P.K. Sahoo$^{2}${\includegraphics[scale=0.06]{Orcid-ID.png}}}\email{pksahoo@hyderabad.bits-pilani.ac.in}

\author{\href{https://orcid.org/0000-0001-9859-1758}Kai Zhou$^{1,4}${\includegraphics[scale=0.06]{Orcid-ID.png}}}\email{zhoukai@cuhk.edu.cn } 

\affiliation{$^1$School of Science and Engineering, The Chinese University of Hong Kong, Shenzhen, Guangdong, 518172, China}

\affiliation{$^2$Department of Mathematics, Birla Institute of Technology and Science, Pilani, Hyderabad Campus, Jawahar Nagar, Kapra Mandal, Medchal District, Telangana 500078, India}

\affiliation{$^3$Chennai Mathematical Institute, H1 SIPCOT IT Park, Siruseri 603103, India}

\affiliation{$^4$School of Artificial Intelligence, The Chinese University of Hong Kong, Shenzhen, Guangdong, 518172, China}

\date{\today}

\begin{abstract}
We investigate neutron star (NS) properties in symmetric teleparallel
$f(Q)$ gravity while consistently accounting for uncertainties in the
dense matter equation of state (EOS). We focus on the linear $f(Q)$
model and perform a joint Bayesian inference of the nuclear matter and
modified gravity parameters using a Taylor expansion EOS around nuclear
saturation density. The analysis incorporates complementary constraints
from nuclear saturation properties, chiral effective field theory
calculations of pure neutron matter, heavy-ion collision, and empirical
nuclear data, the GW170817 tidal deformability posterior, and NICER
mass--radius measurements. The resulting posterior distributions
simultaneously constrain the dense matter and gravity sectors and
quantify their correlations within a common framework. For a canonical
$1.4\,M_\odot$ NS, we obtain
$R_{1.4}=11.28^{+0.35}_{-0.31}\,\mathrm{km}$ and
$\Lambda_{1.4}=151.4^{+44.9}_{-31.2}$, while the maximum mass is
constrained to
$M_{\rm max}=2.22^{+0.29}_{-0.16}\,M_\odot$.
The modified gravity parameters exhibit only weak direct correlations
with most stellar observables, while a pronounced degeneracy remains
within the gravity sector. Our results show that accounting for EOS
uncertainties is essential for quantifying the interplay between
dense matter physics and modified gravity and provide a systematic
framework for confronting symmetric teleparallel gravity with nuclear
and multimessenger NS observations.
\end{abstract}


\maketitle
\section{Introduction}\label{intro}

General relativity (GR) has passed every precision test from laboratory scales to astrophysical and cosmological observations~\citep{will2014}. Even so, open problems such as the late-time accelerated expansion of the Universe and the nature of dark energy~\citep{riess1998, peebles2003} have motivated a broad search for modified theories of gravity. The most familiar route keeps the curvature-based description and replaces the Ricci scalar $R$ in the Einstein-Hilbert action by a general function $f(R)$~\citep{sotiriou2010}, at the cost of fourth-order field equations. Curvature, however, is not the only way to describe gravity. According to the geometrical trinity of gravity~\citep{jimenez2019trinity}, GR can be rewritten equivalently in terms of torsion, as in the teleparallel equivalent of GR~\citep{aldrovandi2013, bahamonde2021}, or in terms of non-metricity, as in the symmetric teleparallel equivalent of GR (STEGR)~\citep{jimenez2018coincident}. In STEGR, the connection is flat and torsion-free, and gravity is carried entirely by the non-metricity tensor $Q_{\alpha\mu\nu} = \nabla_{\alpha} g_{\mu\nu}$. Although these three formulations give the same dynamics as GR, the equivalence is lost once they are extended to $f(R)$, $f(T)$ or $f(Q)$, so each extension can lead to physically distinct predictions~\citep{capozziello2022}. Promoting the non-metricity scalar to an arbitrary function gives $f(Q)$ gravity~\citep{jimenez2018coincident, heisenberg2024review}. Unlike $f(R)$ gravity, $f(Q)$ gravity keeps the field equations for the metric second order. It has been studied widely in cosmology~\citep{jimenez2020cosmology, anagnostopoulos2023} and increasingly for static, spherically symmetric and compact objects~\citep{lin2021, wang/2022, zhao2022covariant}. Neutron stars, with their strong gravitational fields and high central densities, provide a natural setting in which such modifications can be tested beyond cosmological scales.

The detection of gravitational waves (GWs) from the binary neutron star (BNS) merger GW170817 by the Advanced LIGO and Advanced Virgo detectors~\citep{LIGOScientific:2017vwq, acernese2015}, together with its electromagnetic counterpart~\citep{LIGOScientific:2017ync}, opened the era of multimessenger astronomy of neutron stars (NSs). During the inspiral, the tidal field of each star deforms its companion. The size of this deformation, quantified by the tidal deformability $\Lambda$, depends on the internal structure of the star, and it leaves a measurable imprint on the GW phase. The signal therefore carries direct information about the NS equation of state (EOS)~\citep{hinderer2008, oertel2017, chatziioannou2020}. Complementary information comes from the Neutron Star Interior Composition Explorer (NICER). By modelling the X-ray pulses emitted from hot spots on the stellar surface, NICER has provided simultaneous mass and radius estimates for PSR J0030+0451~\citep{riley2019, miller2019}, PSR J0740+6620~\citep{riley2021}, PSR J0437$-$4715~\citep{choudhury2024} and PSR J0614$-$3329~\citep{mauviard2025}. Combining these multimessenger observations has led to increasingly tight constraints on NS properties and on dense matter~\citep{biswas2021}. However, almost all such analyses assume that GR is the correct theory of gravity. This assumption matters because the equilibrium structure of a NS is determined jointly by the EOS and by the theory of gravity. A change in one can be compensated by a change in the other, and the uncertainties from the two sectors currently overlap to a large extent~\citep{biswas2024degeneracy}. The strong gravitational fields inside NSs therefore make it necessary to treat the uncertainties in the EOS and in the gravitational theory simultaneously rather than fixing either one.

The EOS of dense matter, i.e.\ the relation between pressure and energy density, remains the main source of uncertainty in modeling NSs~\citep{ oertel2017}. Near the nuclear saturation density, $n_0 \simeq 0.16~\mathrm{fm}^{-3}$, the EOS is reasonably well constrained. There, the energy per nucleon of asymmetric matter can be expanded in a Taylor series around $n_0$. The expansion coefficients are the nuclear empirical parameters: the binding energy, incompressibility and skewness of symmetric matter, and the magnitude, slope and curvature of the symmetry energy~\citep{margueron2018a}. The lower-order coefficients are fixed fairly well by finite nuclei, whereas the higher-order ones, which control the behaviour at larger densities, remain poorly known. Additional information comes from chiral effective field theory ($\chi$EFT) calculations of pure neutron matter up to about $1$--$2\,n_0$~\citep{hebeler2013, drischler2019}, and from heavy-ion collisions and symmetry-energy measurements at supra-saturation densities~\citep{danielewicz2002, tsang2009}. Beyond these densities the EOS can only be constrained indirectly through astrophysical observations. Bayesian frameworks that combine nuclear-physics inputs with NICER and GW data have therefore become the standard approach in GR~\citep{rutherford2024, brandes2025, imam2025}. In modified gravity, by contrast, NS studies usually adopt a single EOS or a small set of tabulated EOSs, and the gravity parameters are constrained conditional on that choice~\citep{Pradhan:2026wte}. Two problems follow. First, the inferred gravity parameters absorb any mismatch between the chosen EOS and the true one. Second, many realistic EOSs have themselves been calibrated using astrophysical data interpreted within GR. A consistent test of gravity with NSs therefore requires an EOS description anchored in nuclear physics, whose parameters are inferred together with those of the gravity sector~\citep{nobleson2023, biswas2024degeneracy}.

In our previous work~\citep{Pradhan:2026wte}, we performed a Bayesian
analysis of three $f(Q)$ models (linear, logarithmic, and exponential)
using NICER mass--radius measurements, with the EOS fixed to DDME2.
That analysis illustrated the sensitivity of the inferred gravity sector
parameters to the assumed dense matter model. Since the EOS and the
theory of gravity jointly determine the stellar structure, fixing the
EOS can cause part of the model dependence to be absorbed into the
inferred gravity parameters. In addition, many realistic EOSs have
themselves been calibrated using astrophysical information interpreted
within GR. A consistent test of gravity with NSs therefore requires an
EOS description anchored in nuclear physics, whose parameters are
inferred together with those of the gravity sector
~\citep{nobleson2023,biswas2024degeneracy}. More recently, GW170817
tidal information has also been employed to constrain NS configurations
in modified gravity frameworks, including $f(R,T)$ and nonminimally
coupled $f(R,L_m)$ models
~\cite{Lobato:2021ehf,Lobato:2020fxt}. These studies further illustrate
the growing role of multimessenger observations in probing strong-field
gravity with neutron stars.
In this work, we perform a joint Bayesian inference of the dense matter
EOS and the gravity sector within symmetric teleparallel $f(Q)$ gravity.
For the matter sector, we adopt a Taylor expansion EOS around nuclear
saturation density and treat the nuclear matter parameters as free
quantities. For the gravity sector, we consider the linear $f(Q)$ model,
which provides a consistent framework within the static,
spherically symmetric coincident-gauge formulation adopted here. The
analysis combines constraints from nuclear saturation properties,
$\chi$EFT calculations of pure neutron matter, heavy-ion collision and
empirical nuclear data, the GW170817 tidal deformability posterior, and
NICER mass--radius measurements. By sampling the dense matter and
gravity parameters simultaneously, we quantify the correlations and
degeneracies between the two sectors and determine their joint impact on
neutron star properties.

The paper is organized as follows: Sec.~\ref{EOS} introduces the Taylor-expanded dense-matter EOS around nuclear saturation density. Sec.~\ref{sec3} describes the modified Tolman--Oppenheimer--Volkoff (TOV) equations that govern neutron star structure in $f(Q)$ gravity.  The Bayesian framework used for the joint inference of the nuclear and gravity parameters is presented in Sec.~\ref{bay}. In Sec.~\ref{data}, we introduce the multiphysics data sets entering the likelihood, namely the $\chi$EFT calculations of pure neutron matter, heavy-ion collision and empirical nuclear data, the GW170817 tidal-deformability posterior, and the NICER mass--radius measurements. Our results, including the posterior distributions of the nuclear and gravity parameters, their correlations, and the inferred neutron star properties, are discussed in Sec.~\ref{results}. Finally, we summarize our findings and present concluding remarks in Sec.~\ref{sec7}.

\section{Equation of State of Dense Matter}
\label{EOS}

We construct the dense matter equation of state using a Taylor expansion around the nuclear saturation density ($\rho_0$) \cite{Chen:2005ti, Chen:2009wv, Newton:2014iha, Margueron:2017lup, Margueron:2018eob, Imam:2021dbe, Patra:2022yqc}. The relevant nuclear matter parameters are treated as free parameters within the Bayesian analysis, allowing us to explore the density dependence of symmetric nuclear matter and the symmetry energy. Within the parabolic approximation, the energy per nucleon $\varepsilon(\rho,\delta)$, at baryon density $\rho$ and isospin asymmetry $\delta$, is expressed as

\bea
\varepsilon(\rho, \delta) &=&  \varepsilon(\rho,0)+ E_{sym}(\rho)\delta^2
+..., \label{eq:EoS}
\eea
where $\delta = \frac{\rho_n -\rho_p}{\rho}$ is calculated  using the $\beta$-equilibrium and the charge neutrality conditions. The energy per particle for SNM \(\varepsilon(\rho, 0)\) and symmetry energy \(E_{sym}(\rho)\) can be evaluated using nuclear matter parameters, a$_n$ and b$_n$, which are defined as 
\begin{equation}
a_n = (3\rho_0)^n \frac{d^n \varepsilon(\rho, 0)}{d\rho^n} \Bigg|_{\rho = \rho_0}
\end{equation}

where,
\(a_0 = e_0\) (binding energy),
\(a_1 = 0\),
\(a_2 = K_0\) ( incompressibility coefficient),
\(a_3 = Q_0\) (skewness),
\(a_4 = Z_0\) (kurtosis), and

\begin{equation}
b_n = (3\rho_0)^n \frac{d^n E_{sym}(\rho, 0)}{d\rho^n} \Bigg|_{\rho = \rho_0}
\end{equation}

where
\(b_0 = J_0\) (symmetry energy coefficient),
\(b_1 = L_0\) (slope parameter),
\(b_2 = K_{\text{sym0}}\) (curvature parameter),
\(b_3 = Q_{\text{sym0}}\) (skewness),
\(b_4 = Z_{\text{sym0}}\) (kurtosis).

At low densities, the core EOS is smoothly connected to the crust EOS.
We adopt the Baym--Pethick--Sutherland (BPS) EOS~\cite{Baym:1971pw}
for the outer crust up to $\rho=0.0016\rho_0$, while the core EOS
is used above $0.5\rho_0$. In the intermediate density
region, $0.0016\rho_0<\rho<0.5\rho_0$, the inner crust is described
by a polytropic interpolation,

\begin{equation}
\label{bps}
p(\varepsilon)=c_1+c_2\varepsilon^{4/3},
\end{equation}

where the coefficients $c_1$ and $c_2$ are determined by requiring
pressure continuity at the outer-crust--inner-crust and
inner-crust--core boundaries. The resulting crust-core EOS should satisfy three essential conditions: (i)
thermodynamic stability, (ii) causality $(c_s^2 < c^2)$, and (iii)
positive symmetry energy. This complete EOS provides the matter input for the stellar-structure equations in $f(Q)$ gravity, which are introduced in the following section.

\section{Modified TOV Equations in $f(Q)$ Gravity}\label{sec3}

Throughout this work, we adopt the symmetric teleparallel formulation of
gravity, in which the affine connection is flat and torsion-free and the
gravitational dynamics are encoded in the nonmetricity scalar $Q$. Since
the general formalism is well established, we do not reproduce it here and refer the reader to Refs.~\citep{jimenez2018coincident,heisenberg2024review}
for its formulation and to Refs.~\citep{Pradhan:2026wte,wang/2022} for the conventions adopted in the present work. In particular, we work in the coincident gauge, $\Gamma^{\alpha}{}_{\mu\nu}=0$, for which the
nonmetricity scalar is determined entirely by the metric.

\subsection{Field equations for a static, spherically symmetric star}

The interior of a static, spherically symmetric star is described by the line element
\begin{equation}
    ds^{2} = -e^{\nu(r)} dt^{2} + e^{\lambda(r)} dr^{2} + r^{2} d\Omega^{2},
    \label{eq:metric}
\end{equation}
with $d\Omega^{2} = d\theta^{2} + \sin^{2}\theta \, d\phi^{2}$. The two metric potentials $\nu(r)$ and $\lambda(r)$ carry the $tt$ and $rr$ components of the geometry, respectively. The stellar matter is treated as a perfect isotropic fluid \cite{Tolman:1939jz, Oppenheimer:1939ne},
\begin{equation}
    T_{\mu\nu} = \left( \varepsilon + P \right) u_{\mu} u_{\nu} + P g_{\mu\nu},
    \label{eq:fluid}
\end{equation}
where $\varepsilon(r)$ and $P(r)$ are the energy density and pressure, and the four-velocity satisfies $u^{\mu} u_{\mu} = -1$.

For the ansatz~(\ref{eq:metric}), the non-metricity scalar depends only on the radial coordinate,
\begin{equation}
    Q(r) = - \frac{2 e^{-\lambda}}{r} \left( \nu' + \frac{1}{r} \right),
    \label{eq:Q_radial}
\end{equation}
where a prime denotes $d/dr$. Substituting Eqs.~(\ref{eq:metric})--(\ref{eq:Q_radial}) into the $f(Q)$ field equations yields the independent components
\begin{align}
   \kappa  \varepsilon &= \frac{f(Q)}{2}
    - f_Q \left[ Q + \frac{1}{r^{2}}
    + \frac{e^{-\lambda}}{r} \left( \nu' + \lambda' \right) \right],
    \label{eq:eom_density} \\[4pt]
   \kappa P &= - \frac{f(Q)}{2} + f_Q \left( Q + \frac{1}{r^{2}} \right),
    \label{eq:eom_radial} \\[4pt]
   \kappa P &= - \frac{f(Q)}{2}
    + f_Q \left\{ \frac{Q}{2}
    - e^{-\lambda} \left[ \frac{\nu''}{2}
    + \left( \frac{\nu'}{4} + \frac{1}{2r} \right) \left( \nu' - \lambda' \right)
    \right] \right\},
    \label{eq:eom_tangential} \\[4pt]
    0 &= \frac{\cot\theta}{2}\, Q'\, f_{QQ},
    \label{eq:offdiag}
\end{align}
where $\kappa = 8\pi G / c^{4}$, $f_Q \equiv df/dQ$ and $f_{QQ} \equiv d^{2}f/dQ^{2}$.

Equation~(\ref{eq:offdiag}) is the off-diagonal $r\theta$ component, it has no counterpart in GR. Its $\cot\theta$ factor cannot be cancelled by anything else in a spherically symmetric configuration, so the equation must be satisfied by the remaining factors. For a non-linear model, $f_{QQ} \neq 0$, and the condition can only be met if $Q' = 0$, that is, if the non-metricity scalar is constant throughout the star. This is a direct consequence of the coincident gauge: fixing the connection to zero exhausts the gauge freedom, and the residual field equation then constrains the allowed configurations rather than being satisfied identically. The same point has been raised recently in analyses that keep the connection dynamical~\citep{heisenberg2025neutronstarsfmathbbqgravity, Ayuso:2025vkc}.

\subsection{Stellar structure equations}

The conservation of the energy-momentum tensor gives the usual hydrostatic equilibrium condition,
\begin{equation}
    P' + \frac{\nu'}{2} \left( \varepsilon + P \right) = 0.
    \label{eq:tov_hydro}
\end{equation}
Together with Eqs.~(\ref{eq:eom_density})--(\ref{eq:eom_tangential}), this closes the system. Eliminating $\varepsilon$ and $P$ in favour of the metric functions, the structure of the star is governed by

\begin{align}
    \lambda' &= \frac{- \kappa r \left( \varepsilon + P \right) e^{\lambda} - f_Q\nu'}{f_Q},
    \label{eq:tov_lambda} \\[4pt]
    \nu'' &= \frac{1}{2 r f_Q}
    \Big[ 2 r Q f_Q e^{\lambda}
    - \left( r \nu' + 2 \right) \left( \nu' - \lambda' \right) f_Q \nonumber \\
    &\qquad\quad - 2 \left( 2 \kappa P + f(Q) \right) r e^{\lambda} \Big],
    \label{eq:tov_nu}
\end{align}
Equations~(\ref{eq:tov_hydro})--(\ref{eq:tov_nu}) are the modified TOV equations of $f(Q)$ gravity. Once a functional form of $f(Q)$ and an EOS relating $P$ to $\varepsilon$ are specified, they are integrated outward from the centre subject to the regularity conditions
\begin{equation}
    \lambda(0) = 0, \qquad \nu'(0) = 0,
    \label{eq:bc}
\end{equation}
until the pressure vanishes, $P(R) = 0$, which defines the stellar surface at radius $R$. The exterior is vacuum, and for $T_{\mu\nu} = 0$ Eqs.~(\ref{eq:eom_density}) and (\ref{eq:eom_radial}) give
\begin{equation}
    \nu'(r) + \lambda'(r) = 0,
    \label{eq:vacuum}
\end{equation}
so the interior solution is matched to the Schwarzschild geometry by requiring $e^{\nu(R)} = e^{-\lambda(R)}$ at the surface.

In the limit $f(Q) = -Q$ one has $f_Q = -1$ and $f_{QQ} = 0$, so Eq.~(\ref{eq:offdiag}) is satisfied identically and Eqs.~(\ref{eq:eom_density})--(\ref{eq:eom_tangential}) reduce to the standard Einstein field equations. Departures from GR are therefore controlled entirely by the chosen functional form of $f(Q)$. In this work we adopt the linear model,
\begin{eqnarray}
f(Q)=Q_m \bigg[\alpha+\beta \Big( \frac{Q}{Q_m} \Big)  \bigg],
    \label{eq:lin_model}
\end{eqnarray}

The linear model is the only case in which the coincident gauge adopted here is internally consistent with $f_{QQ}=0$ the off-diagonal equation~(\ref{eq:offdiag}) is satisfied identically, so the field equations impose no further restriction on $Q$ and the spherically symmetric ansatz~(\ref{eq:metric}) can be used without over-constraining the solutions. This consistency comes at a definite price, which is worth making explicit. Substituting $f_Q=\beta$ and $f_{QQ}=0$ into the covariant field equations, the derivative term proportional to $\partial_\lambda Q$ vanishes and the remaining bracket reduces to a constant, $f_Q Q - f = -\alpha Q_m$, leaving
\begin{equation}
    \beta\, \mathring{G}_{\mu\nu} 
    +\frac{1}{2}\alpha Q_m\, g_{\mu\nu} = -\kappa T_{\mu\nu}.
    \label{eq:linear_fieldeq}
\end{equation}
It is important to note that, in the linear model, $\alpha$ and $Q_m$
enter the field equations only through the product $\alpha Q_m$.
Consequently, NS observables cannot independently determine
these two parameters within the present formulation, leading to an
intrinsic degeneracy between them. The individual marginalized
posteriors of $\alpha$ and $Q_m$ should therefore be interpreted with
this parameter degeneracy in mind, while the combination $\alpha Q_m$
is the quantity directly relevant to the stellar-structure equations.
The non-metricity sector therefore enters only through two constants: $\beta$ rescales the gravitational coupling, $G_{\rm eff} = G/|\beta|$, while $\alpha Q_m$ acts as an effective cosmological term in the stellar interior. General relativity is recovered for $\alpha = 0$ and $\beta = -1$. The connection does not appear in Eq.~(\ref{eq:linear_fieldeq}) at all, which is precisely why the gauge choice is harmless here, and it also means that the model does not probe non-metricity as a dynamical degree of freedom.

What the present analysis constrains is consequently a well-posed but specific question. Because $G_{\rm eff}$ is a single constant, common to the weak-field and strong-field regimes, variations in $\beta$ can be partially degenerate with changes in the EOS in stellar-structure observables. In particular, a smaller $|\beta|$ corresponds to a larger $G_{\rm eff}$ and therefore to a stronger effective gravitational coupling, which can be compensated by a stiffer EOS providing greater pressure support. The two effects can thus produce similar changes in the mass--radius relation, giving rise to an EOS--gravity degeneracy. Sampling $\{\alpha, \beta, Q_m\}$ jointly with the nuclear matter parameters measures directly how much of the inferred stellar structure can be reassigned from the matter sector to the gravitational sector, and how strongly the multiphysics constraints of Sec.~\ref{data} break that reassignment.

\section{Bayesian Analysis}\label{bay}

We employ Bayesian inference to constrain simultaneously the parameters
associated with the dense-matter EOS and the linear $f(Q)$ gravity
model. We denote the complete set of model parameters by $\boldsymbol{\theta}$, which includes the nuclear matter parameters entering the Taylor-expanded EOS together with the parameters characterizing the $f(Q)$ model. For a given dataset $\mathcal{D}$, Bayes' theorem gives the posterior distribution as
\begin{equation}
P(\boldsymbol{\theta}|\mathcal{D},H)
=
\frac{
\mathcal{L}(\mathcal{D}|\boldsymbol{\theta},H)
\,\pi(\boldsymbol{\theta}|H)
}{
\mathcal{Z}(\mathcal{D}|H)
},
\label{eq:bayes}
\end{equation}
where $\mathcal{L}(\mathcal{D}|\boldsymbol{\theta},H)$ denotes the likelihood of the data for a parameter set $\boldsymbol{\theta}$, while
$\pi(\boldsymbol{\theta}|H)$ represents the corresponding prior distribution. The quantity $\mathcal{Z}(\mathcal{D}|H)$ is the Bayesian evidence and provides the normalization of the posterior. The resulting posterior distributions are used to determine the allowed parameter ranges and correlations between the dense-matter and modified-gravity sectors.

\subsection{Likelihood}\label{likelihood}

For a given parameter set \(\boldsymbol{\theta}\), the likelihood measures the consistency of the model predictions with the adopted experimental and astrophysical constraints. Under the assumption that the datasets are independent, the joint likelihood is obtained by multiplying the individual contributions.

For an experimental quantity $D_{\rm exp}$ reported with a Gaussian
uncertainty $\sigma$, we adopt
\begin{equation}
\mathcal{L}_{\rm exp}
=
\frac{1}{\sqrt{2\pi\sigma^2}}
\exp\left[
-\frac{\left(D(\boldsymbol{\theta})-D_{\rm exp}\right)^2}
{2\sigma^2}
\right],
\label{eq:gaussian-like}
\end{equation}
Here, \(D(\boldsymbol{\theta})\) represents the theoretical value corresponding to a given parameter set. We use this Gaussian form for the nuclear and terrestrial observables characterized by measured central values and uncertainties, including the saturation properties, pure neutron matter, and heavy-ion collision constraints.

For gravitational-wave observations, the likelihood depends on the component masses and their corresponding tidal deformabilities. For a given $\boldsymbol{\theta}$, the EOS and the $f(Q)$ gravity parameters uniquely determine the relation $\Lambda(M)$. The GW likelihood can therefore be written as
\begin{align}
\mathcal{L}_{\rm GW}(\boldsymbol{\theta})
&=
P(d_{\rm GW}|\boldsymbol{\theta}) \nonumber\\
&=
\int dm_1\,dm_2\,
P(m_1,m_2|\boldsymbol{\theta}) \nonumber\\
&\quad\times
P\left[
d_{\rm GW}\,\big|
m_1,m_2,
\Lambda_1(m_1,\boldsymbol{\theta}),
\Lambda_2(m_2,\boldsymbol{\theta})
\right].
\label{eq:GW-like}
\end{align}
The mass distribution is taken to be uniform within the adopted mass interval,
\begin{equation}
P(m|\boldsymbol{\theta})
=
\begin{cases}
\dfrac{1}{M_u-M_l}, & M_l\leq m\leq M_u,\\[2mm]
0, & {\rm otherwise}.
\end{cases}
\label{eq:mass-prior}
\end{equation}
Here $M_l$ and $M_u$ denote the lower and upper limits of the mass range
considered for the GW event.

The NICER measurements provide joint posterior information on the neutron-star
mass and radius. For each sampled parameter set, the corresponding
mass--radius relation is obtained by solving the modified stellar-structure
equations. The NICER likelihood is then evaluated as
\begin{align}
\mathcal{L}_{\rm NICER}(\boldsymbol{\theta})
&=
P(d_{\rm X-ray}|\boldsymbol{\theta}) \nonumber\\
&=
\int_{M_l}^{M_u} dm\,
P(m|\boldsymbol{\theta})\,
P\left[
d_{\rm X-ray}\,\big|
m,R(m,\boldsymbol{\theta})
\right].
\label{eq:nicer-like}
\end{align}
For the X-ray likelihood, we take $M_l=1\,M_{\odot}$, while $M_u$ is determined
by the maximum stable neutron-star mass predicted for the corresponding
parameter set. The likelihoods from the different NICER pulsars are multiplied
assuming that the measurements are statistically independent.

The complete likelihood employed in the present analysis is therefore written
schematically as
\begin{equation}
\mathcal{L}_{\rm tot}
= \mathcal{L}_{exp}
\,
\mathcal{L}_{\chi{\rm EFT}}
\,
\mathcal{L}_{\rm HIC}
\,
\mathcal{L}_{\rm GW}
\,
\mathcal{L}_{\rm NICER},
\label{eq:total-like}
\end{equation}
where $\mathcal{L}_{\chi{\rm EFT}}$ accounts for the microscopic
pure-neutron-matter information, and $\mathcal{L}_{\rm HIC}$ incorporates the
constraints obtained from heavy-ion collision experiments. The last two terms
contain the gravitational-wave and NICER information, respectively. In this
way, terrestrial nuclear information and astrophysical observations are
included within a common Bayesian framework to constrain both the EOS and the
parameters of linear $f(Q)$ gravity.

\section{Multiphysics Data}\label{data}

In the present analysis, all nuclear, terrestrial, and astrophysical constraints are incorporated simultaneously within a common Bayesian framework. The adopted inputs include microscopic pure-neutron-matter calculations, empirical and heavy-ion collision constraints on nuclear matter, the GW170817 tidal-deformability posterior, and the available NICER mass--radius measurements. These complementary datasets probe different density regimes and therefore provide joint constraints on both the dense-matter EOS and the parameters of the linear $f(Q)$ model.

\subsection{$\chi$EFT}

At sub-saturation and moderately suprasaturation densities, chiral effective field theory provides a controlled description of neutron-rich matter. In particular, $N^{3}\mathrm{LO}$ calculations based on $NN+3N$ interactions constrained by nucleon--nucleon scattering, few-body observables, and nuclear saturation properties provide reliable predictions for pure neutron matter~\cite{Hebeler:2013nza, Drischler:2021kxf, Drischler:2020hwi}. We use the pure-neutron-matter pressure, $P_{\rm PNM}$, obtained from the $N^{3}\mathrm{LO}$ calculations up to a density of $0.34~\mathrm{fm}^{-3}$~\cite{Drischler:2020hwi, Drischler:2021kxf}.

\subsection{Heavy-Ion Collision}

We include empirical constraints on the symmetry energy $J(\rho)$, symmetry-energy pressure $P_{\rm sym}$, and symmetric nuclear matter pressure $P_{\rm SNM}$. These constraints are inferred from nuclear masses, the neutron-skin thickness of $^{208}\mathrm{Pb}$, dipole polarizability, isobaric analog states, and heavy-ion collision measurements over the density interval $0.03$--$0.32~\mathrm{fm}^{-3}$, following the compilation given in Table~I of Refs.~\cite{Tsang:2023vhh, Imam:2024gfh, Imam:2024xqg}.

\subsection{GW170817}

The tidal response of neutron stars in binary mergers provides an important probe of the high-density EOS. For GW170817, whose component masses lie approximately in the range $1.17$--$1.6~M_\odot$, gravitational-wave observations provide posterior information on the component tidal deformabilities~\cite{Abbott:2017vwq}. We use the publicly available GW170817 posterior samples released by the LIGO--Virgo Collaboration\footnote{LVK Collaboration,
\href{https://dcc.ligo.org/LIGO-P1800115/public}
{https://dcc.ligo.org/LIGO-P1800115/public}} as an astrophysical constraint in the Bayesian analysis. We note that these posterior samples were inferred using waveform models constructed within GR. In the present work, they are therefore employed as phenomenological constraints on the tidal deformabilities predicted within the adopted $f(Q)$ framework. A fully self-consistent treatment would require the binary dynamics and waveform model to be formulated directly in the corresponding modified-gravity theory.

\subsection{NICER}

Mass--radius measurements from NICER provide direct information on neutron-star structure through X-ray pulse-profile modeling. We include the publicly available posterior distributions for PSR J0030+0451 obtained independently by Riley {\sl et al.}~\cite{Riley:2019yda}~\footnote{~\href{https://zenodo.org/records/8239000}{https://zenodo.org/records/8239000}} and Miller {\sl et al.}~\cite{Miller:2019cac}~\footnote{~\href{https://zenodo.org/record/3473466}{https://zenodo.org/record/3473466}}, together with the corresponding measurements of PSR J0740+6620 from Riley {\sl et al.}~\cite{Riley:2021pdl}~\footnote{~\href{https://zenodo.org/records/4697625}{https://zenodo.org/records/4697625}} and Miller {\sl et al.}~\cite{Miller:2021qha}~\footnote{~\href{https://zenodo.org/records/4670689}{https://zenodo.org/records/4670689}}. We further include the more recent NICER posteriors for PSR J0437-4715 from Choudhury {\sl et al.}~\cite{Choudhury:2024xbk}~\footnote{~\href{https://zenodo.org/records/13766753}{https://zenodo.org/records/13766753}} and PSR J0614-3329 from Mauviard {\sl et al.}~\cite{Mauviard:2025dmd}~\footnote{~\href{https://zenodo.org/records/15603406}{https://zenodo.org/records/15603406}}. Together, these measurements provide coverage over a broad region of the neutron-star mass--radius plane and are incorporated jointly with the nuclear and gravitational-wave constraints.

\section{Results and Discussion}\label{results}

In this section, we present the results of the joint Bayesian inference of the dense-matter EOS and the linear \(f(Q)\) gravity model. The Taylor-expansion EOS is combined with the modified stellar-structure equations, and the nuclear matter and \(f(Q)\) parameters are constrained simultaneously using the complete set of nuclear, heavy-ion, gravitational wave, and NICER data introduced in Sec.~\ref{data}. We first discuss the prior, posterior distributions and parameter correlations, and then examine the resulting constraints on the nuclear matter properties, the \(\beta\)-equilibrated EOS, and NS properties, including the mass--radius relation, tidal deformability, and maximum mass. 

\subsection{Priors and Posterior distributions of parameters}

The posterior distributions of the nuclear matter parameters (NMPs) and the linear \(f(Q)\) model parameters are obtained by simultaneously imposing the nuclear, theoretical, and astrophysical constraints described in Sec.~\ref{data}. These include the basic properties of nuclear matter around saturation density, the pure-neutron-matter pressure predicted by \(N^3\)LO chiral effective field theory, empirical information from heavy-ion collision experiments, and the GW170817 and NICER observations. The adopted prior distributions for the NMPs and the \(f(Q)\) parameters are summarized in Table~\ref{tab1:prior}. The saturation density \(\rho_0\) and binding energy per nucleon \(e_0\) are assigned Gaussian priors, while uniform priors are adopted for the remaining NMPs and the modified-gravity parameters \(\alpha\), \(\beta\), and \(Q_m\).

\begin{table}[h!]
\caption{\label{tab1:prior}Prior distributions of the NMPs. e$_0$ and $\rho_0$ are assumed as Gaussian distributions with a mean($\mu$) and standard deviation ($\sigma$). All other NMPs are assumed to be uniformly distributed between a minimum (min.) and a maximum (max.) values as listed in the units of MeV. In below cells, we have listed the prior distributions of $f(Q)$ model parameters. } 
  \centering
\setlength{\tabcolsep}{1.pt}
\renewcommand{\arraystretch}{1.4}
  \begin{ruledtabular}
  \begin{tabular}{ccccccccccc}
&{$\rho_0$} & {e$_0$} &{K$_0$} & {Q$_0$} & {Z$_0$} &{J$_0$} & {L$_0$} &{K$_{\rm sym,0}$} & {Q$_{\rm sym,0}$}& {Z$_{\rm sym,0}$}\\ [1.3ex]
 \hline
 min./$\mu$ & 0.160 & -16.0 & 150 & -1500 & -5000 & 25 & 20 & -400 & -1500 &-5000\\[1.3ex] 
 max./$\sigma$ & 0.005 & 0.30 & 330 & 1500 & 5000 & 40 & 120 & 400 & 1500 &5000\\[1.3ex] 
 \hline \hline
 &{$\alpha$} & {$\beta$} &{$Q_m$} &  &  &  &  & & & \\ [1.3ex]
 \hline
 min. & -15.0 & -5.0 & -1.0 &  &  &  &  &  & &\\[1.3ex] 
 max. & 15.0 & 5.0 & 1.0 &  &  &  &  &  &  &\\[1.3ex] 
  \end{tabular}
  \end{ruledtabular}
\end{table}

\begin{figure*}
    \centering
    \includegraphics[width=\linewidth]{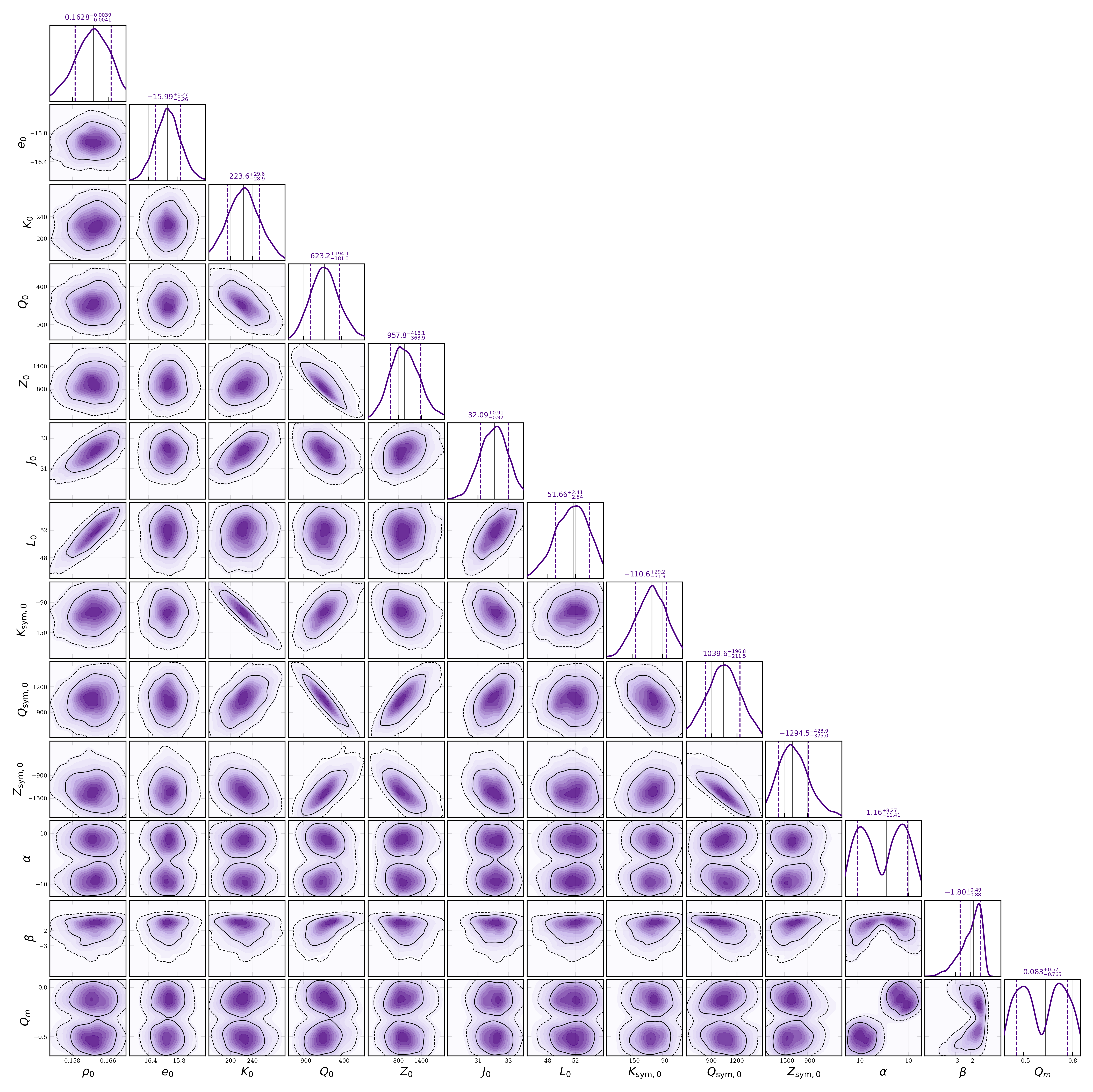}
    \caption{The marginalized posterior distributions of the NMPs and $f(Q)$ model parameters, obtained through Bayesian inference.}
    \label{fig1}
\end{figure*}

\begin{figure*}
   \centering 
 \includegraphics[width=\textwidth]{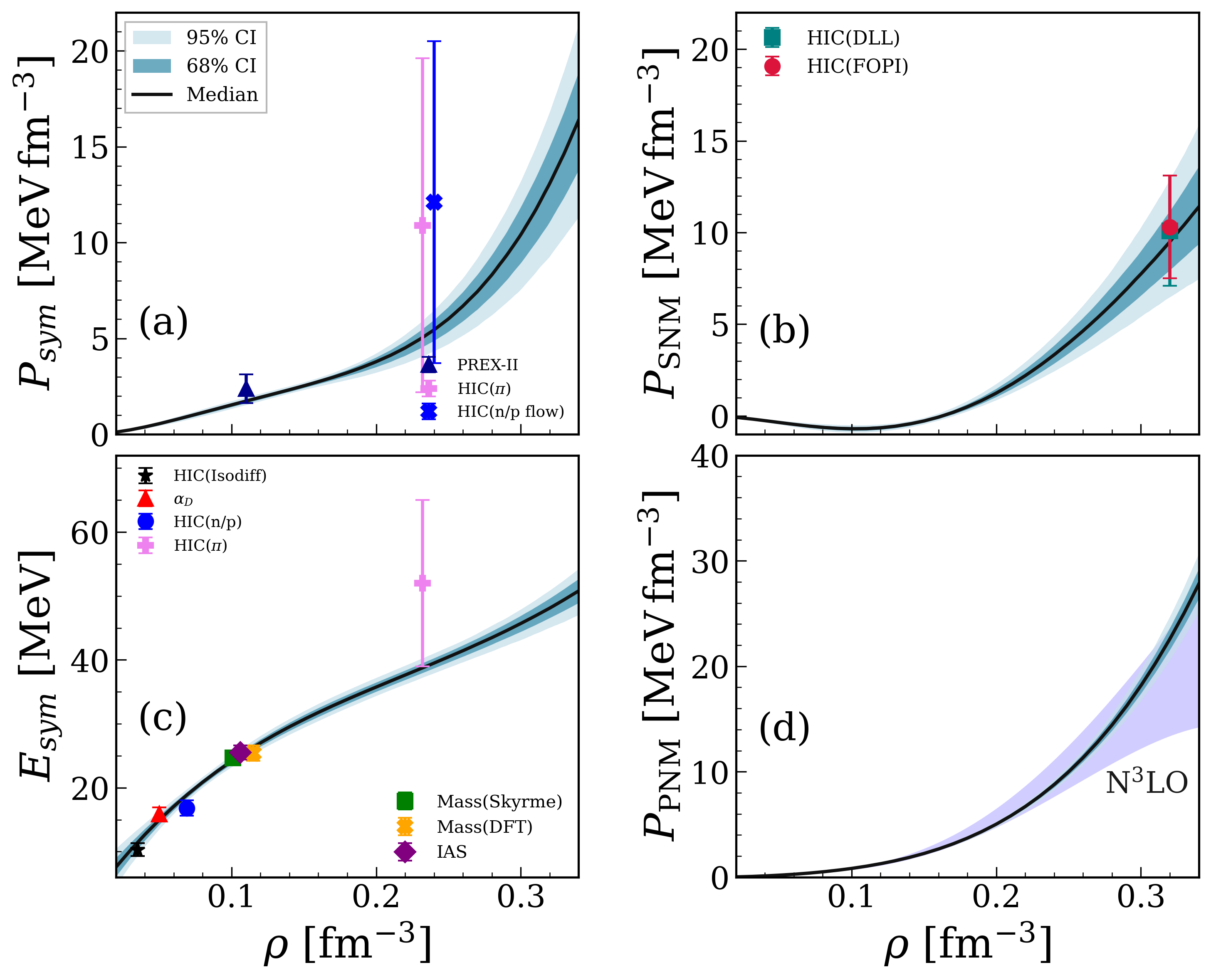}
    \caption{The 95\% and 68 \% credible interval plots for (a) the pressure of the symmetry energy, $P_{\rm sym}$, (b) the pressure of SNM, $P_{\rm SNM}$, (c) the symmetry energy, $E_{sym}(\rho)$, and (d) the pressure of PNM, $P_{\rm PNM}$, as a function of the baryon density $\rho$, evaluated using the posterior distributions as listed in Fig.~\ref{fig1}. The corresponding experimental data incorporated into the Bayesian framework are also shown.}
    \label{fig2}
\end{figure*}

The joint posterior distribution of the nuclear matter parameters (NMPs) and the \(f(Q)\)-gravity parameters is obtained from the product of the likelihood and prior distributions, as defined in Eq.~\ref{eq:bayes}. The marginalized posterior distribution of each parameter is obtained by integrating the joint posterior over all remaining model parameters. A posterior distribution that is appreciably narrower than its corresponding prior indicates that the parameter is constrained by the nuclear-theory and astrophysical data entering the likelihood. The one- and two-dimensional marginalized posterior distributions obtained for the Taylor-expanded EOS combined with the \(f(Q)\) model are shown in Fig.~\ref{fig1}. The median values and the corresponding \(68\%\) credible intervals are reported above the diagonal panels, while the off-diagonal panels display the joint \(68\%\) and \(95\%\) credible regions.

The posterior distributions of several nuclear matter parameters (NMPs) are significantly narrower than their adopted priors, indicating that the combined nuclear and astrophysical constraints provide substantial information on the density dependence of the EOS. At the \(68\%\) credible level, we obtained $L_0=51.66^{+2.41}_{-2.54}\ {\rm MeV}$,$ 
Q_0=-623.22^{+194.15}_{-181.25}\ {\rm MeV}$, $Z_0=957.79^{+416.13}_{-363.85}\ {\rm MeV}$, $K_{{\rm sym},0}=-110.62^{+29.16}_{-31.86}\ {\rm MeV}$.
The posteriors of \(K_0\), \(J_0\), \(Q_{{\rm sym},0}\), and \(Z_{{\rm sym},0}\) are also appreciably constrained relative to their broad uniform priors, whereas \(e_0=-15.99^{+0.27}_{-0.26}\) MeV remains largely prior dominated. For the linear \(f(Q)\) model, the modified-gravity parameters are inferred as $\alpha=1.16^{+8.27}_{-11.41}$, $
\beta=-1.80^{+0.49}_{-0.88}$, $Q_m=0.083^{+0.571}_{-0.765}$,
with the posterior favoring \(\beta<0\), while the \(68\%\) credible intervals of both \(\alpha\) and \(Q_m\) include zero. The marginalized posteriors of $\alpha$ and $Q_m$ exhibit pronounced
bimodal features, reflecting the intrinsic degeneracy associated with
the fact that these parameters enter the stellar-structure equations
only through the combination $\alpha Q_m$.

Several pronounced correlations are visible in the joint posterior distributions. The strongest anticorrelations occur between \(Q_0\) and \(Q_{{\rm sym},0}\), and between \(K_0\) and \(K_{{\rm sym},0}\), with \(r\simeq-0.92\), while \(\rho_0\) and \(L_0\) exhibit a strong positive correlation, \(r=0.85\). Other notable correlations include \(\alpha\)–\(Q_m\) (\(r=0.80\)), \(Q_{{\rm sym},0}\)–\(Z_{{\rm sym},0}\) (\(r=-0.76\)), and \(Q_0\)–\(Z_0\) (\(r=-0.73\)). Moderate correlations, with \(|r|\simeq0.59\!-\!0.69\), are also found for \(Z_0\)–\(Q_{{\rm sym},0}\), \(\rho_0\)–\(J_0\), \(Q_0\)–\(Z_{{\rm sym},0}\), \(J_0\)–\(L_0\), \(Z_0\)–\(Z_{{\rm sym},0}\), and \(K_0\)–\(J_0\). The strong correlation between \(\alpha\) and \(Q_m\) reflects the degeneracy associated with the constant contribution \(\alpha Q_m\) in the linear \(f(Q)\) model. In contrast, the modified-gravity parameters are only weakly correlated with most NMPs, apart from moderate correlations of \(\beta\) with \(Q_0\) (\(r=0.47\)) and \(Q_{{\rm sym},0}\) (\(r=-0.42\)).

\begin{figure*}
    \centering
  \includegraphics[width=\textwidth]{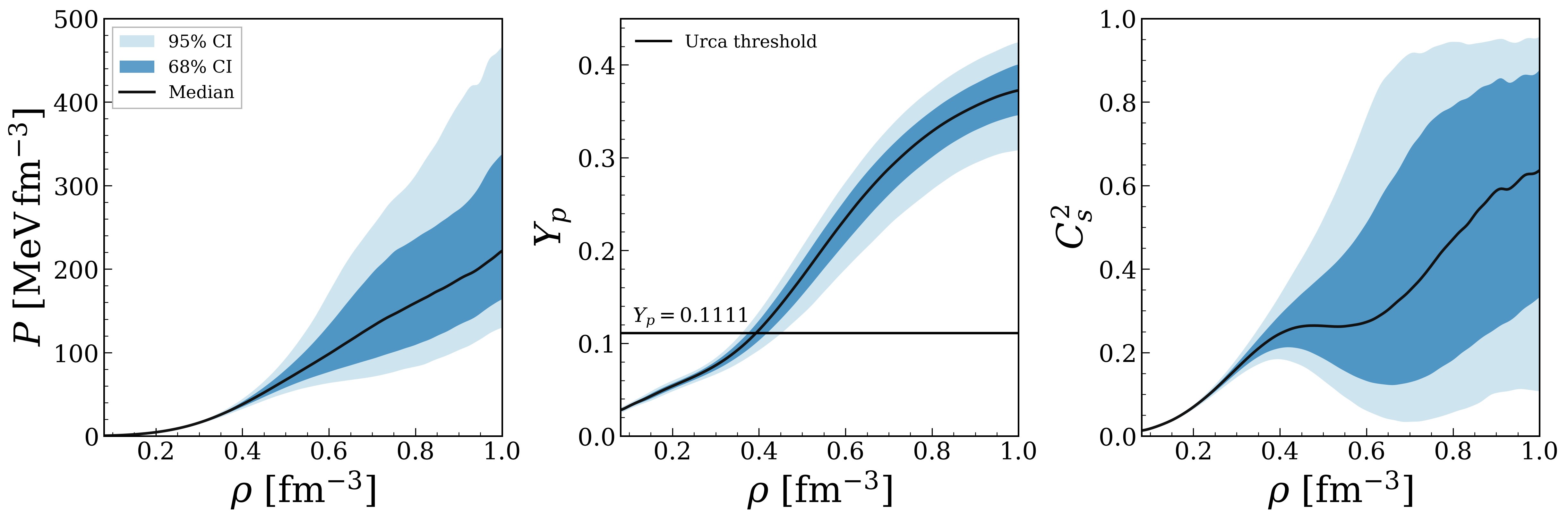}
    \caption{The 95$\%$ and 68\% credible interval posterior distributions of the pressure of $\beta$-equilibrated matter , $P$ (left), the proton fraction , $Y_{\rm p}$ (middle), and the squared sound speed, $c_s^2$ (right), as functions of the baryon density $\rho$. The solid horizontal line indicates the nominal direct-Urca threshold \(Y_p=1/9\).}
    \label{fig3}
\end{figure*}

\subsection{Posterior distributions of the EOS and NS properties}

Using the posterior distributions of the model parameters shown in
Fig.~\ref{fig1}, we reconstruct the density dependence of the nuclear and
$\beta$-equilibrated matter properties. Figure~\ref{fig2} shows the 95\%
and 68\% credible intervals for (a) the symmetry-energy pressure,
$P_{\rm sym}$, (b) the pressure of symmetric nuclear matter,
$P_{\rm SNM}$, (c) the symmetry energy, $E_{\rm sym}(\rho)$, and
(d) the pressure of pure neutron matter, $P_{\rm PNM}$, as functions of
the baryon density $\rho$. The posterior bands are consistent with the
experimental, empirical, and $\chi$EFT constraints included in the
analysis over the density region in which these data are available.
The uncertainties remain relatively narrow around and moderately above
the saturation density, while they gradually broaden toward higher
densities, reflecting the weaker terrestrial constraints in this regime.

Figure~\ref{fig3} presents the corresponding posterior distributions of
the pressure of $\beta$-equilibrated matter, proton fraction $Y_p$, and
squared speed of sound $c_s^2$. The pressure increases smoothly with
density, as required for a thermodynamically stable EOS, while its
uncertainty becomes appreciably larger at supranuclear densities. The
proton fraction increases monotonically and crosses the nominal
direct-Urca threshold, $Y_p=1/9\simeq0.111$, at approximately
$\rho\simeq0.4~{\rm fm}^{-3}$ for the median EOS. This suggests that
sufficiently massive NSs may allow the direct Urca process, subject to
their central densities and detailed particle composition. The squared
sound speed increases with density, exhibits a relatively mild variation
at intermediate densities, and rises again in the high-density regime.
The median prediction remains well below the causal limit,
$c_s^2=1$, throughout the density range considered, while the credible
intervals broaden substantially toward higher densities. Overall, these
results show that the low- and intermediate-density behavior of the EOS
is comparatively well constrained, whereas appreciable uncertainty
remains in the supranuclear-density regime relevant to the cores of
massive NSs.

\begin{figure*}
    \centering
    \includegraphics[width=\textwidth]{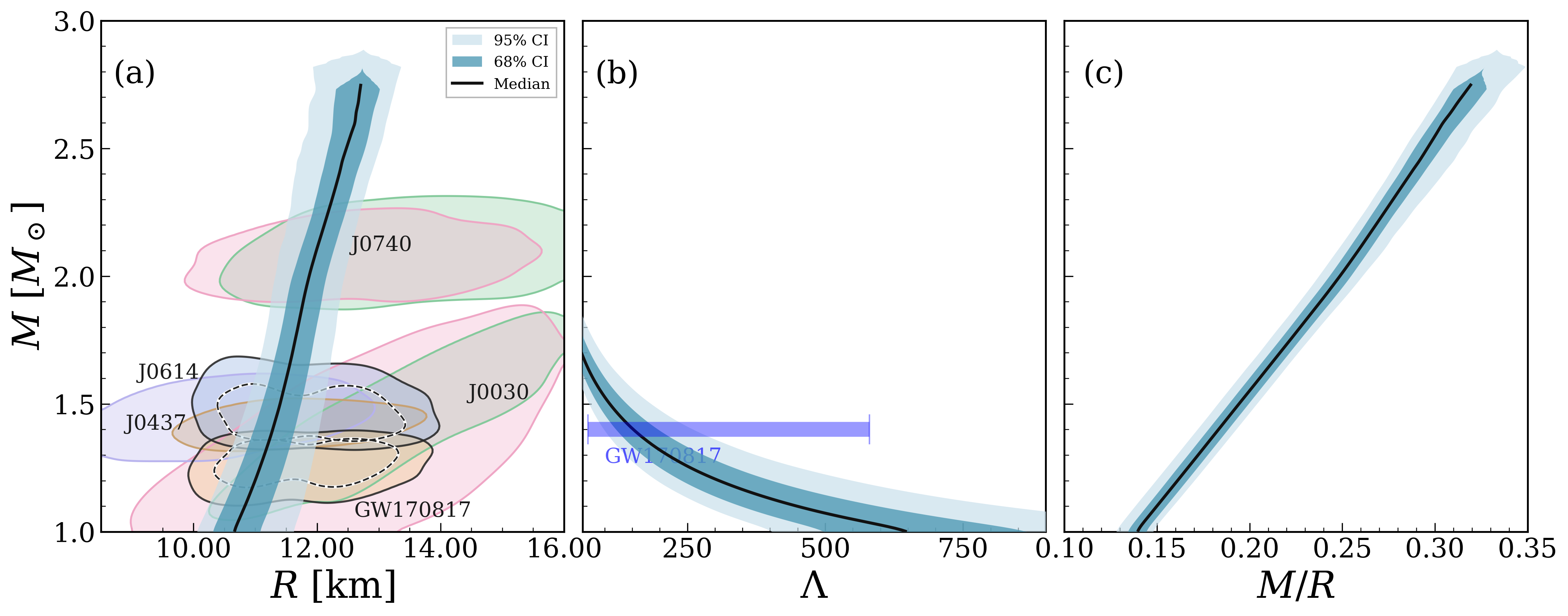}
    \caption{ The 95$\%$ and 68 \% credible interval distributions for the radius, R(km) (left panel) and tidal deformability, $\Lambda$ (middle panel) , and compactness (right panel) as a function of neutron star mass, $M(M_\odot)$. The astrophysical observations incorporated in the Bayesian framework are also shown.}
    \label{fig4}
\end{figure*}

\begin{figure}
    \centering
    \includegraphics[width=\linewidth]{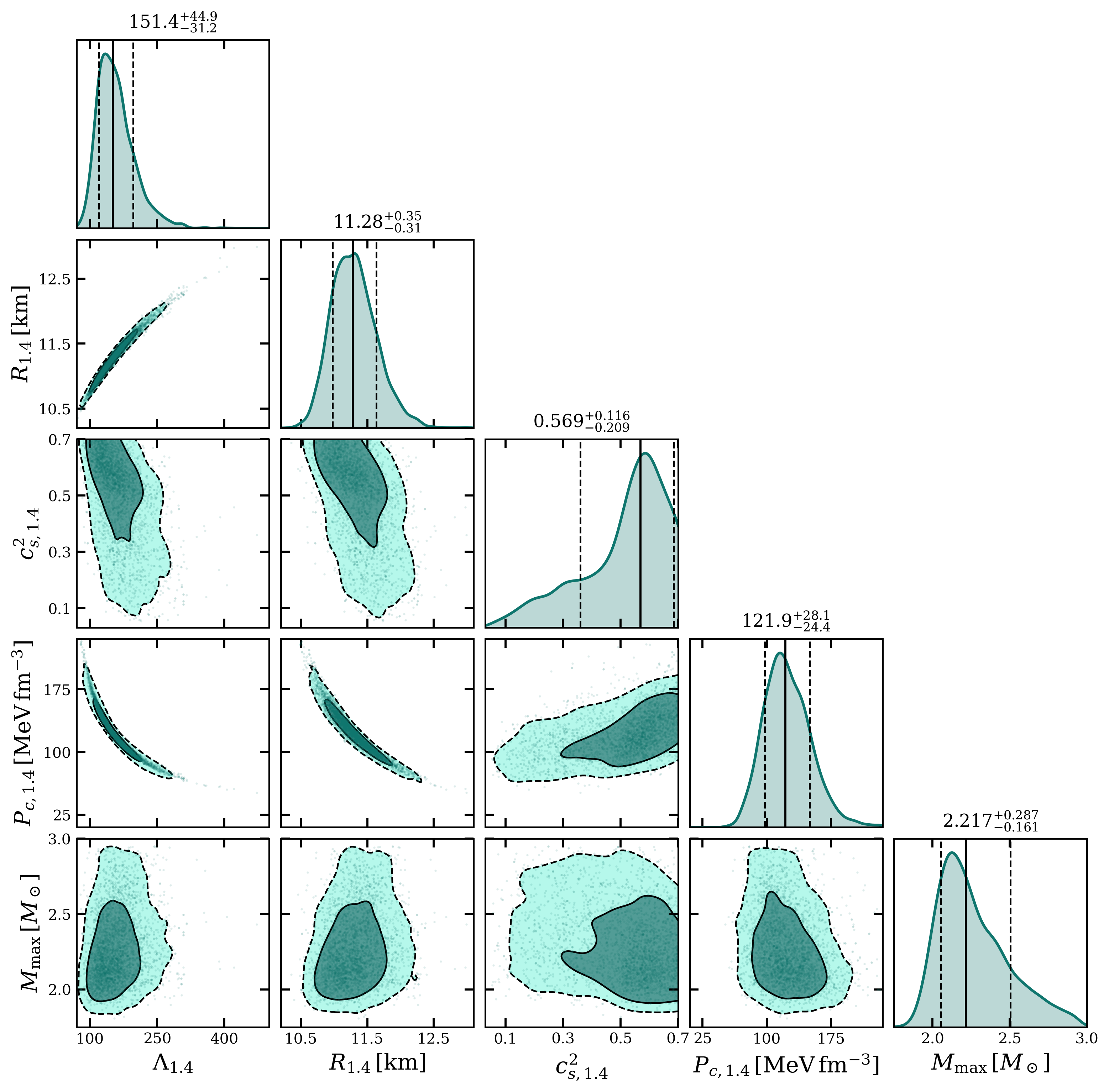}
    \caption{The marginalized posterior distributions of the NS properties, obtained through Bayesian inference.}
    \label{fig5}
\end{figure}

\begin{figure*}
    \centering
    \includegraphics[width=\linewidth]{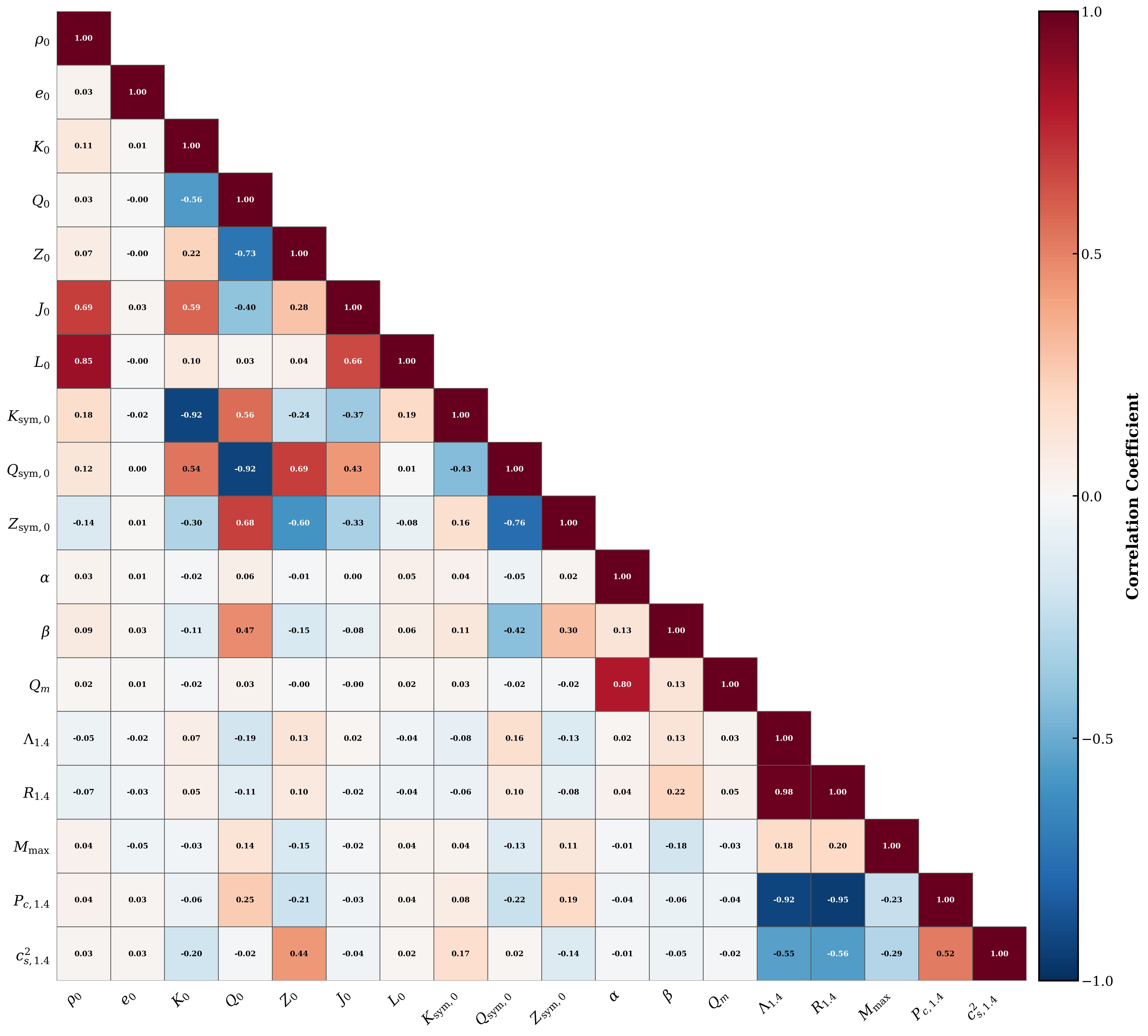}
    \caption{Pearson's correlation matrix plots between the model parameters and selected NS properties.}
    \label{fig6}
\end{figure*}

The corresponding constraints on the macroscopic NS properties are
shown in Fig.~\ref{fig4}. The left panel displays the 68\% and 95\%
credible regions of the mass--radius relation together with the
mass--radius constraints from the NICER observations and GW170817.
The posterior favors a relatively narrow range of radii over a broad
interval of stellar masses and remains compatible with the observational
regions included in the Bayesian analysis. For a canonical
$1.4\,M_\odot$ NS, we obtained
$R_{1.4}=11.28^{+0.35}_{-0.31}$ km. The middle panel shows the tidal
deformability as a function of stellar mass. As expected, $\Lambda$
decreases rapidly with increasing mass as the star becomes progressively
more compact. The inferred value at $1.4\,M_\odot$ is
$\Lambda_{1.4}=151.4^{+44.9}_{-31.2}$, consistent with the tidal
deformability constraint from GW170817. The right panel presents the
mass--compactness relation, where
$C=M/R$ increases systematically with stellar mass. Overall, the
posterior bands show that the combined nuclear and astrophysical
constraints restrict the stellar radius, tidal response, and compactness
within the linear $f(Q)$ framework.

To further quantify these results, Fig.~\ref{fig5} shows the marginalized
posterior distributions and correlations among selected NS properties,
namely $\Lambda_{1.4}$, $R_{1.4}$, $c_{s,1.4}^{2}$,
$p_{c,1.4}$, and $M_{\rm max}$. In addition to the constraints on
$\Lambda_{1.4}$ and $R_{1.4}$ quoted above, we obtain
$c_{s,1.4}^{2}=0.569^{+0.116}_{-0.209}$ and
$p_{c,1.4}=121.9^{+28.1}_{-24.4}\ {\rm MeV\,fm^{-3}}$.
The maximum mass is constrained to
$M_{\rm max}=2.217^{+0.287}_{-0.161}\,M_\odot$, indicating that the
posterior EOSs are compatible with the existence of massive neutron
stars above the $2\,M_\odot$ scale.

Several correlations among these quantities are also evident from
Fig.~\ref{fig5}. A strong positive correlation is found between
$R_{1.4}$ and $\Lambda_{1.4}$, reflecting the strong sensitivity of
the tidal deformability to the stellar radius. In contrast, the
central pressure $p_{c,1.4}$ is strongly anticorrelated with both
$R_{1.4}$ and $\Lambda_{1.4}$, with smaller radii and tidal
deformabilities being associated with larger central pressures.
The central squared sound speed $c_{s,1.4}^{2}$ shows broader
anticorrelations with $R_{1.4}$ and $\Lambda_{1.4}$ and a positive
association with $p_{c,1.4}$. These correlations illustrate how the
properties of dense matter are reflected in the macroscopic observables
of neutron stars. By comparison, the correlations involving
$M_{\rm max}$ are considerably broader, consistent with its greater
sensitivity to the high-density behavior of the EOS beyond the densities
predominantly probed by a $1.4\,M_\odot$ star.

\begin{figure*}
    \centering
    \includegraphics[width=\linewidth]{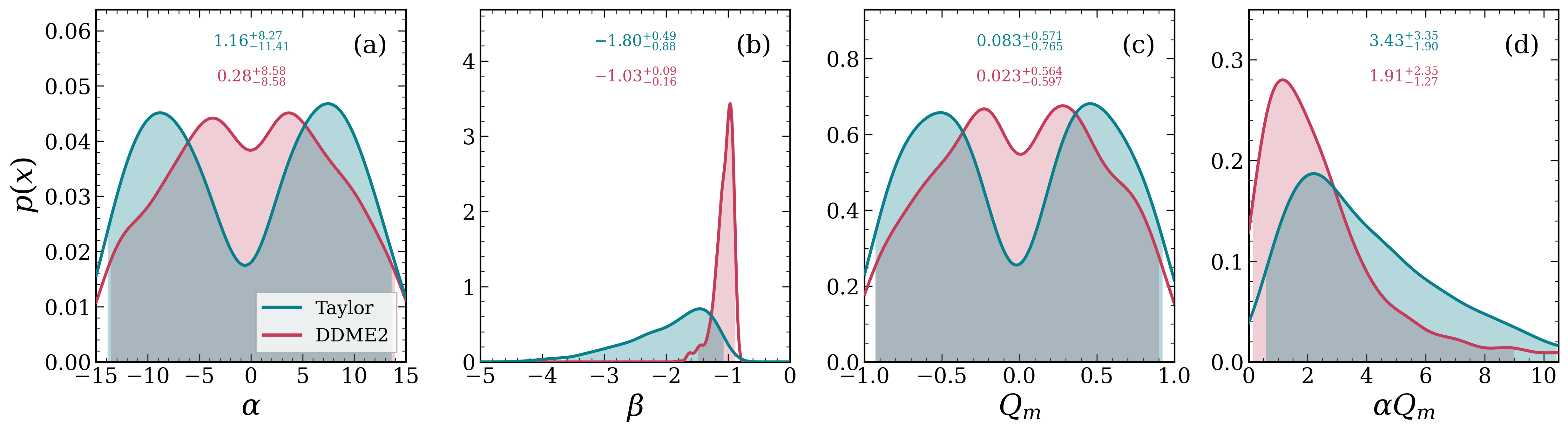}
    \caption{ Comparison of the one-dimensional marginalized posterior distributions of the linear $f(Q)$ parameters $\alpha$, $\beta$, and $Q_m$, together with the derived combination $\alpha Q_m$,  obtained in the present Taylor-expansion EOS analysis and in the previous fixed-DDME2 analysis~\cite{Pradhan:2026wte}. The quoted values denote the median and corresponding 68\% credible intervals.}
    \label{fig7}
\end{figure*}

Finally, to examine the interplay between the dense matter sector,
the modified gravity parameters, and the resulting NS properties, we
evaluate the Pearson correlation coefficients among all sampled
parameters and selected NS observables. The corresponding correlation
matrix is shown in Fig.~\ref{fig6}. Several characteristic correlation
patterns are evident among the NMPs. The strongest anticorrelations are
found between $K_0$ and $K_{\rm sym,0}$ and between $Q_0$ and
$Q_{\rm sym,0}$, both with $r\simeq-0.92$. A strong anticorrelation is
also observed between $Q_{\rm sym,0}$ and $Z_{\rm sym,0}$
($r=-0.76$), while $Q_0$ and $Z_0$ are anticorrelated with
$r=-0.73$. The saturation density $\rho_0$ shows strong positive
correlations with $L_0$ ($r=0.85$) and $J_0$ ($r=0.69$).
These correlations indicate that the combined constraints restrict
specific combinations of NMPs rather than acting independently on each
parameter, reflecting their coupled role in determining the density
dependence of the EOS.

The modified gravity sector exhibits a different correlation pattern.
The strongest correlation is between $\alpha$ and $Q_m$, with
$r=0.80$. This correlation reflects the intrinsic parameter degeneracy
of the linear $f(Q)$ model, since $\alpha$ and $Q_m$ enter the
stellar-structure equations only through the combination
$\alpha Q_m$. Consequently, their correlated joint posterior and the
bimodal features of the corresponding one-dimensional marginalized
posteriors should be interpreted as manifestations of this
non-identifiability rather than as evidence for distinct physical
branches. In this sense, the combination $\alpha Q_m$ is the quantity
directly constrained by the NS properties, whereas $\alpha$
and $Q_m$ individually remain only weakly identifiable.

The parameter $\beta$ shows moderate correlations with $Q_0$
($r=0.47$) and $Q_{\rm sym,0}$ ($r=-0.42$), while its correlations
with the remaining NMPs are generally weak. Direct correlations between
the modified gravity parameters and the NS properties are also small.
In particular, $\alpha$ and $Q_m$ show negligible correlations with
$\Lambda_{1.4}$, $R_{1.4}$, $M_{\rm max}$, $P_{c,1.4}$, and
$c_{s,1.4}^{2}$, whereas $\beta$ exhibits only a weak positive
correlation with $R_{1.4}$ ($r=0.22$) and weak correlations with the
other stellar quantities. This behavior suggests that, within the
linear $f(Q)$ framework, the stellar properties are predominantly
shaped by the dense matter sector, with only a limited direct
sensitivity to the modified gravity parameters over the posterior
region explored.

Among the NS properties, the strongest correlation is found between
$\Lambda_{1.4}$ and $R_{1.4}$, with $r=0.98$. The central pressure
$p_{c,1.4}$ is strongly anticorrelated with both $R_{1.4}$
($r=-0.95$) and $\Lambda_{1.4}$ ($r=-0.92$). The central squared
sound speed $c_{s,1.4}^{2}$ is moderately anticorrelated with
$R_{1.4}$ ($r=-0.56$) and $\Lambda_{1.4}$ ($r=-0.55$), while it is
positively correlated with $p_{c,1.4}$ ($r=0.52$). In contrast,
$M_{\rm max}$ shows only weak correlations with the canonical-star
observables, with $r=0.20$ for $R_{1.4}$, $r=0.18$ for
$\Lambda_{1.4}$, $r=-0.23$ for $p_{c,1.4}$, and $r=-0.29$ for
$c_{s,1.4}^{2}$. The comparatively weak correlations involving
$M_{\rm max}$ are consistent with its greater sensitivity to the
high-density EOS, which is less tightly constrained than the density
region predominantly probed by a $1.4\,M_\odot$ star.

It is also instructive to compare the present results with our previous
fixed-DDME2 analysis~\cite{Pradhan:2026wte}, in which the linear,
logarithmic, and exponential $f(Q)$ models were investigated separately.
For the linear model, the earlier analysis showed a strong correlation
between $R_{1.4}$ and $\Lambda_{1.4}$, while $M_{\rm max}$ remained only
weakly correlated with the other stellar observables. The gravity-sector
parameters also exhibited a moderate correlation between $\alpha$ and
the characteristic nonmetricity scale, whereas $\beta$ was comparatively
well constrained around the GR-like regime. These qualitative features
persist in the present analysis when the EOS parameters are allowed to
vary simultaneously with the $f(Q)$ parameters. In particular,
$R_{1.4}$ and $\Lambda_{1.4}$ remain very strongly correlated
($r=0.98$), while $M_{\rm max}$ continues to show only weak correlations
with the canonical-star observables. The correlation between $\alpha$
and $Q_m$ becomes more pronounced ($r=0.80$), reflecting the degeneracy
associated with their combined contribution to the linear $f(Q)$ model.
At the same time, the direct correlations of the modified gravity
parameters with the NS properties remain generally weak.

To further examine the impact of EOS uncertainties on the gravity sector, Fig.~\ref{fig7} compares the one-dimensional marginalized posterior distributions of the linear $f(Q)$ parameters with those obtained in our previous fixed-DDME2 analysis~\cite{Pradhan:2026wte}. The posteriors of $\alpha$ and $Q_m$ remain broad in both analyses, while the present calculation exhibits a more pronounced bimodal structure, consistent with the strong $\alpha$--$Q_m$ correlation found in Fig.~\ref{fig6}. This behavior is a direct consequence of the intrinsic degeneracy between these parameters. As follows from Eq.~\ref{eq:lin_model}, the linear model can be written as
$f(Q)=\alpha Q_m+\beta Q$, so that $\alpha$ and $Q_m$ enter the
stellar-structure equations only through their product $\alpha Q_m$.

This interpretation is further supported by the posterior distribution
of $\alpha Q_m$ shown in the fourth panel of Fig.~\ref{fig7}. In
contrast to the pronounced bimodal structures of the individual
$\alpha$ and $Q_m$ posteriors, the distribution of their product is
single-peaked and considerably smoother. We obtain
$\alpha Q_m=3.43^{+3.35}_{-1.90}$ in the present analysis, compared
with $1.91^{+2.35}_{-1.27}$ in the previous fixed-DDME2 calculation.
The substantial overlap between the two distributions indicates that
the bimodal behavior of $\alpha$ and $Q_m$ does not represent distinct
physical branches, but instead reflects their non-identifiability when
treated as separate parameters. The combination $\alpha Q_m$ therefore
provides a more direct measure of the gravity-sector contribution and
a more meaningful quantity for comparison between the two analyses.

The most significant difference between the two analyses remains in
$\beta$. In the fixed-DDME2 analysis, $\beta$ was tightly localized
around $\beta=-1.03^{+0.09}_{-0.16}$, whereas the present joint
inference gives $\beta=-1.80^{+0.49}_{-0.88}$ with a substantially
broader posterior. This broadening is consistent with an increased
matter--gravity degeneracy when the dense matter EOS is allowed to vary
simultaneously with the $f(Q)$ parameters. Since the two analyses differ
not only in their EOS treatment but also in the nuclear and astrophysical information entering the likelihood, the comparison should be regarded as illustrative rather than as a controlled EOS-model test.

\section{Conclusion}\label{sec7}

In this work, we have performed a joint Bayesian analysis of the dense
matter equation of state and neutron star structure within the framework
of symmetric teleparallel $f(Q)$ gravity. For the matter sector, we
adopted a Taylor expansion EOS around the nuclear saturation density,
treating the nuclear matter parameters as free quantities. For the
gravity sector, we considered the linear $f(Q)$ model and constrained
its parameters simultaneously with the EOS parameters. The Bayesian
inference combines complementary information from nuclear saturation
properties, chiral effective field theory calculations of pure neutron
matter, empirical and heavy-ion collision constraints, the GW170817
tidal deformability posterior, and the available NICER mass--radius
measurements. In this way, the dense matter and modified-gravity
parameters are inferred consistently within a common framework.

The resulting NS properties are compatible with the astrophysical
constraints included in the analysis. For a canonical
$1.4\,M_\odot$ star, we obtain
$R_{1.4}=11.28^{+0.35}_{-0.31}\ {\rm km}$ and
$\Lambda_{1.4}=151.4^{+44.9}_{-31.2}$, while the maximum NS mass is
constrained to
$M_{\rm max}=2.217^{+0.287}_{-0.161}\,M_\odot$.
The corresponding central properties are
$c_{s,1.4}^{2}=0.569^{+0.116}_{-0.209}$ and
$p_{c,1.4}=121.9^{+28.1}_{-24.4}\ {\rm MeV\,fm^{-3}}$.
These results show that the posterior EOSs comfortably support neutron
stars above the $2\,M_\odot$ mass scale. The inferred correlations
further reveal a strong relation between $R_{1.4}$ and
$\Lambda_{1.4}$, whereas $M_{\rm max}$ remains only weakly correlated
with the canonical star observables. Within the gravity sector,
$\alpha$ and $Q_m$ exhibit a pronounced correlation, while the direct
correlations of the modified gravity parameters with the stellar
observables remain generally weak.

The present analysis therefore provides a more complete picture of the
EOS--gravity interplay in neutron stars. Allowing the dense matter EOS
to vary simultaneously with the $f(Q)$ parameters makes it possible to
quantify how the observational constraints are distributed between the
matter and gravity sectors. Canonical star observables such as
$R_{1.4}$ and $\Lambda_{1.4}$ primarily probe the EOS at intermediate
supranuclear densities, whereas $M_{\rm max}$ remains more sensitive to
the less constrained high-density sector. Our results show that reliable
constraints on modified gravity from neutron stars require the matter
and gravity sectors to be inferred simultaneously, rather than fixing one
sector while constraining the other. This joint treatment constitutes the
main advance of the present work and provides a systematic framework for
confronting symmetric teleparallel gravity with nuclear and neutron star
observations.

\section{Acknowledgements}
NP and KZ acknowledge support by the CUHK-Shenzhen University development fund under grant No. UDF01003041 and UDF03003041, and Shenzhen Peacock fund under No. 2023TC0179, Ministry of Science and Technology of China under Grant No. 2024YFA1611004, NSFC fund under No. 92570117.  PKS acknowledges IUCAA, Pune, India for providing support through the visiting Associateship program.

\bibliographystyle{apsrev4-1}
\bibliography{ref}

\section*{\Large Appendix: Comparison with the Previous Fixed-DDME2 Analysis} \label{appendix}

For completeness, we compare the stellar predictions obtained in the
present joint Taylor expansion EOS analysis with those from our previous
fixed-DDME2 calculation~\cite{Pradhan:2026wte} for the linear $f(Q)$
model. Since the two analyses differ in both the EOS treatment and the
information entering the likelihood, the comparison is intended to
illustrate the resulting changes in the posterior predictions rather
than to provide a controlled EOS model comparison.

\begin{figure*}
    \centering
    \includegraphics[width=\linewidth]{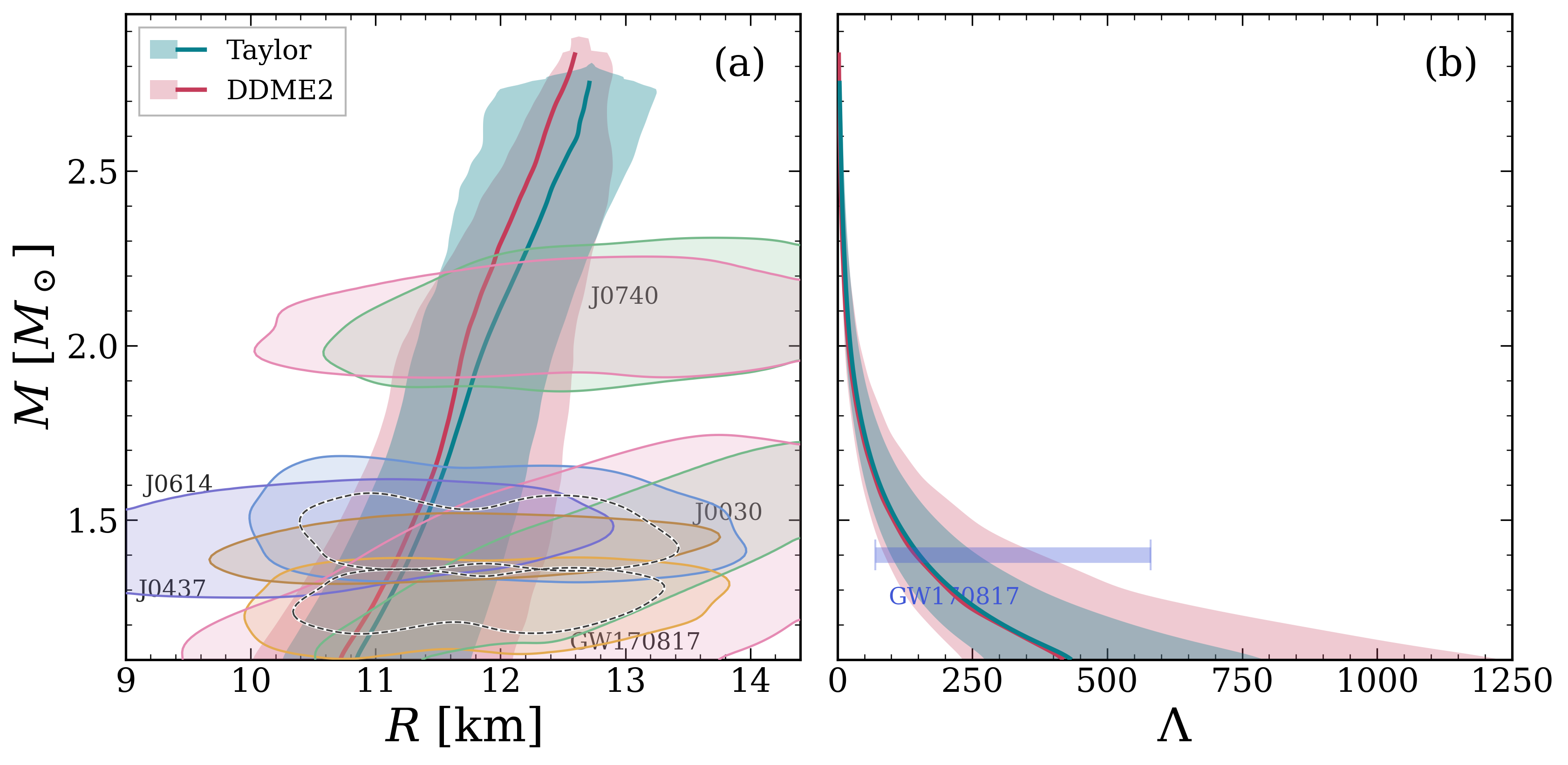}
    \caption{ Similarly to Fig.~\ref{fig7}, but showing the 95\% credible regions of the neutron-star mass--radius relation (left panel) and mass--tidal-deformability relation (right panel).}
    \label{fig8}
\end{figure*}

\begin{figure*}
    \centering
    \includegraphics[width=\linewidth]{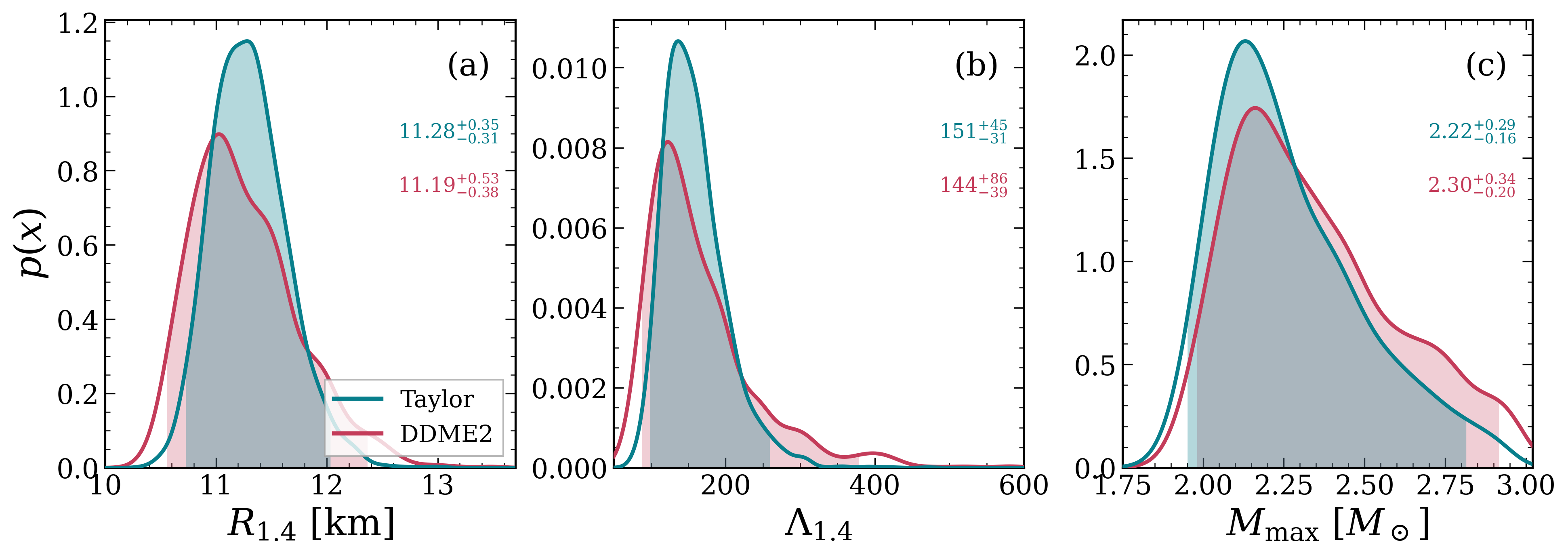}
    \caption{
Similarly to Fig.~\ref{fig7}, but showing the one-dimensional
marginalized posterior distributions of the neutron-star properties
$R_{1.4}$, $\Lambda_{1.4}$, and $M_{\rm max}$. The quoted values denote
the medians and corresponding 68\% credible intervals.}\label{fig9}
\end{figure*}

Figure~\ref{fig8} compares the 95\% credible regions of the
mass--radius and mass--tidal-deformability relations. The present
Taylor expansion EOS analysis yields visibly narrower posterior bands
over most of the relevant stellar-mass range, while remaining broadly
consistent with the corresponding fixed-DDME2 predictions. The reduced
spread is consistent with the additional nuclear and multimessenger
constraints restricting the allowed dense matter parameter space in the
joint inference. The corresponding one-dimensional marginalized posterior distributions
of $R_{1.4}$, $\Lambda_{1.4}$, and $M_{\rm max}$ are shown in
Fig.~\ref{fig9}. The median values of $R_{1.4}$ and $\Lambda_{1.4}$
remain close to those obtained in the fixed-DDME2 analysis, whereas their
credible intervals are noticeably reduced in the present calculation.
The maximum-mass distribution also becomes narrower and shifts toward a
slightly lower median value. These results are consistent with the
additional EOS constraints reducing the range of stellar configurations
allowed by the joint analysis, including the high-density sector relevant
to $M_{\rm max}$.

\end{document}